\documentclass[rnote]{aa} 
\usepackage{graphicx}
\usepackage{txfonts}
\usepackage{natbib}

\def\cal{1RXS~J141256.0+792204}

\begin{document}
   \title{A Strong Upper Limit on the Pulsed Radio Luminosity of the Compact Object \cal}

   \author{J.~W.~T. Hessels\inst{1}
          \and
          B.~W. Stappers\inst{1,2}
	  \and
	  R.~E. Rutledge\inst{3}
	  \and
	  D.~B. Fox\inst{4}
	  \and
	  A.~H. Shevchuk\inst{4}
	  }

   \offprints{J.~W.~T. Hessels}

   \institute{Astronomical Institute ``Anton Pannekoek'', University of
              Amsterdam, Kruislaan 403, 1098 SJ Amsterdam, The Netherlands\\
              \email{jhessels@science.uva.nl}
         \and
	      Stichting ASTRON, Postbus 2, 7990 AA Dwingeloo, The Netherlands\\
              \email{stappers@astron.nl}
         \and
              Department of Physics, McGill University, 3600 University,
	      H3A 2T8 Montreal, QC, Canada\\
              \email{rutledge@physics.mcgill.ca}
	 \and
	      Department of Astronomy \& Astrophysics, 525 Davey Laboratory, 
	      Pennsylvania State University, University Park, PA 16802, USA\\
	      \email{dfox@astro.psu.edu, ahs148@psu.edu}
        }

   \date{Received July 21, 2007}

 
  \abstract
  {The {\it ROSAT} X-ray source \cal\ has recently been identified as a
  likely compact object whose properties suggest it could be a
  very nearby radio millisecond pulsar at $d = 80 - 260$\,pc.}
  {We investigated this hypothesis by searching for radio
  pulsations using the Westerbork Synthesis Radio Telescope.}
  {We observed \cal\ at 385 and 1380\,MHz, recording at high time and
  frequency resolution in order to maintain sensitivity to millisecond
  pulsations.  These data were searched both for dispersed single pulses 
  and using Fourier techniques sensitive to constant and orbitally modulated periodicities.}
  {No radio pulsations were detected in these observations, resulting in
  pulsed radio luminosity limits of $L_{400}^{\rm max} \approx 0.3 (d/250 {\rm
  pc})^2$\,mJy kpc$^2$ and $L_{1400}^{\rm max} \approx 0.03 (d/250 {\rm
  pc})^2$\,mJy kpc$^2$ at 400 and 1400\,MHz respectively.}
  {The lack of detectable radio pulsations from \cal\ brings into question its
  identification as a nearby radio pulsar, though, because the pulsar could
  be beamed away from us, this hypothesis cannot be strictly ruled out.}


  \titlerunning{\cal\ Radio Luminosity Limit}

  \maketitle
%

\section{Introduction}

Recently, \citet*{rfs07}, hereafter RFS07, have identified the X-ray source
\cal\ (from the {\it ROSAT} All-Sky Survey Bright Source Catalog) as having
an X-ray to optical flux ratio $F_{\rm X}(0.1-2.4 {\rm keV})/F_{\rm V} >
8700$ ($3\sigma$).  Such a high ratio is strong evidence that \cal, dubbed
``Calvera'' by RFS07, is a compact object.  RFS07 consider several specific
source classes to explain \cal's X-ray spectrum and luminosity: an X-ray dim
isolated neutron star (INS), an anomalous X-ray pulsar (AXP), a compact
central object (CCO), or a nearby radio pulsar.  Based on careful comparison
of \cal's properties with the canonical features displayed by these
different classes of neutron star, RFS07 conclude that the most likely
explanation is that \cal\ is a very nearby radio pulsar ($d = 80-260$\,pc)
similar to the radio millisecond pulsars (MSPs) residing in the globular
cluster 47~Tuc.

We have investigated this interpretation by conducting sensitive searches
for radio pulsations using the Westerbork Synthesis Radio Telescope (WSRT)
in the Netherlands.  No pulsations were found by these searches, and we use
these non-detections to place strong limits on the pulsed radio luminosity
of \cal.  These limits bring into question the interpretation of this object
as a nearby radio pulsar.  


\section{Observations and analysis}


We observed \cal\ a total of 4 times with WSRT at central observing
frequencies of 385 and 1380\,MHz and with integration times of 0.5 or 1\,hr
(Table~\ref{obs.tab}). All observations were centered on J2000 coordinates
RA = 14:12:55.51, DEC = +79:22:06.7 ($l = 118.3^{\circ}$, $b =
37.0^{\circ}$). WSRT was used in its tied array mode with 14 25-m dishes
phased together to form a $\sim 43/12^{\prime \prime}$ wide beam at
385/1380\,MHz, with a gain of 1.2\,K/Jy. The most precise X-ray position
available, which is derived from {\it Chandra}/HRC-I observations of the
source taken by RFS07, falls well within the WSRT beam.  We recorded data
using the PuMa I pulsar backend \citep{vkh+02}, which provided 0.16/0.32-MHz
channels, 10/80\,MHz of bandwidth, and 102.4/102.4\,$\mu$s sampling at
385/1380\,MHz.  We made multiple, multi-frequency observations because of
potential scintillation.  If \cal\ is indeed very close, then it likely has
a very low DM ($\lesssim 10$\,pc cm$^{-3}$), and could strongly scintillate.
The 385-MHz data also gave high sensitivity in the event that \cal\ is a very
steep spectrum source. The resulting intra-channel dispersive smearing in
these configurations, scaled to a dispersion measure (DM) of 25\,pc
cm$^{-3}$, is $t_{\rm DM} = 0.6 ({\rm DM} / 25\,{\rm pc~ cm}^{-3})$\,ms at
385\,MHz and merely $t_{\rm DM} = 22 ({\rm DM} / 25\,{\rm pc~
cm}^{-3})$\,$\mu$s for the 1380-MHz observations.  At the high Galactic
latitude of the source, interstellar scattering is unlikely to add
significant additional smearing. The resolution of the 1380-MHz data was
more than adequate to be sensitive to an MSP with spin period $P_{\rm spin}
\gtrsim 0.7$\,ms and DM up to the highest trial DM we searched ($100$\,pc
cm$^{-3}$). In the 385-MHz data, we were sensitive to millisecond
periodicities as well, as long as the DM was less than $\sim 25$\,pc
cm$^{-3}$.  

\begin{table}
\begin{minipage}[t]{\columnwidth}
\caption{Summary of Observations.}
\label{obs.tab}
\centering
\renewcommand{\footnoterule}{}  
\begin{tabular}{cccccc}
\hline 
Epoch    & Int. Time & Freq. & Band.\footnote{Total usable bandwidth / channel
bandwidth.} & Samp. & $S_{\rm max}$\footnote{For the
observation at 385\,MHz this is the upper limit on 400-MHz flux density
$S_{400}$.  For the other three observations at 1380\,MHz, this is the upper
limit on the 1400-MHz flux density $S_{1400}$.} \\
(MJD)    & (s)       & (MHz)     & (MHz)     & $\mu$s  & (mJy) \\
\hline
54217.98 & 1800      & 385       & 10 / 0.16 & 102.4   & 4   \\
54218.01 & 1800      & 1380      & 60 / 0.32 & 102.4   & 0.4 \\
54218.89 & 1800      & 1380      & 60 / 0.32 & 102.4   & 0.4 \\
54245.74 & 3600      & 1380      & 60 / 0.32 & 102.4   & 0.3 \\
\hline
\end{tabular}
\end{minipage}
\end{table}

The DM and spin period of this source are unknown and thus we must search
over these parameters.  To search over DM space, we constructed trial
timeseries dedispersed in the range DM = $0-100$\,pc cm$^{-3}$, with a
spacing of 0.2/1\,pc cm$^{-3}$ for the 385/1380\,MHz data.  Integrating to
the edge of the Galaxy, the maximum predicted DM in the direction of \cal\
is 45\,pc cm$^{-3}$\citep*[from the NE2001 free electron density model of
the Galaxy,][]{cl02}. Thus, even if it is distant, it is very unlikely that
\cal\ has a DM outside our search range. Radio frequency interference (RFI) was
removed from the data both by eliminating some of the spectral channels
before dedispersion and by removing known interference peaks, or ``birdies'',
in the Fourier power spectrum. Each of these timeseries was then searched
for periodicities using a Fourier-based technique and also checked for
strong single pulses.  As the majority of MSPs are in binaries, we employed
searches that are sensitive to orbitally modulated periodicities by
encorporating a further search over frequency drift in the Fourier domain.
Overlapping sections of the timeseries, each 15-min in length, were also
searched to improve sensitivity to the shortest orbital periods. To perform
these searches, we used the package PRESTO\footnote{Available at
http://www.cv.nrao.edu/$\sim$sransom/presto} \citep*[see][for a description
of some of the search techniques included in this package]{rem02}.  Overall,
our search method was very similar, both in terms of the techniques and
specific software used, to that presented in greater detail in e.g.
\citet{hrs+07}.  

The resulting candidate lists were examined by eye and automatically sifted
in order to identify periodicities that occur at several trial DMs, peak in
signal strength at non-zero DM, and occur in multiple observations.  We
folded the raw data at the periodicity, acceleration, and DM of potentially
interesting candidates in order to further investigate whether they
constituted likely astronomical signals or just RFI.  As the predicted DM
for a pulsar in this direction at $d < 260$\,pc is only $< 3$\,pc cm$^{-3}$,
we paid special attention to very-low-DM candidates, which would normally be
dismissed automatically as RFI, carefully checking to see whether they could
plausibly be pulsar signals.  These low-DM candidates were all at the period
of known WSRT RFI birdies, or failed to show up in both the 385 and 1380-MHz
data.

\section{Results and discussion}

No plausible astronomical radio pulsations or bright, dispersed single
pulses were detected in any of our observations.  Using the radiometer
equation modified for pulsar signals \citep{dtws85}, we can place limits on
\cal's flux density in these observations (Table~\ref{obs.tab}). We find
that \cal\ has a maximum flux density at 400\,MHz $S_{400}^{\rm max} \approx
4$\,mJy and a maximum flux density at 1400\,MHz $S_{1400}^{\rm max} \approx
0.3$\,mJy, where the fractional uncertainty on these limits is roughly 50\%.
RFS07 argue that \cal\ may be a radio pulsar with a nearby distance $d =
80-260$\,pc.  Assuming the pulsar is isolated, we were sensitive to spin
periods encompassing the observed range for radio pulsars ($P_{\rm spin} \sim
1$\,ms$-10$\,s), with a factor of roughly $2-5$ degredation in sensitivity
due to red-noise at the longest periods. Using an assumed distance $d =
250$\,pc, we can convert our flux density limits to pseudo luminosity ($L
\equiv Sd^2$) limits.  We find $L_{400}^{\rm max} \approx 0.3 (d/250 {\rm
pc})^2$\,mJy kpc$^2$ and $L_{1400}^{\rm max} \approx 0.03 (d/250 {\rm
pc})^2$\,mJy kpc$^2$.  A simple check of the ATNF pulsar
catalog\footnote{Available at
http://www.atnf.csiro.au/research/pulsar/psrcat} \citep{mhth05} shows that
$\lesssim 1$\% of the known pulsars have a luminosity below these limits.
This suggests that if \cal\ is a radio pulsar, then it is either especially
weak, not beamed towards the Earth, or significantly further away than
250\,pc.

Assuming \cal\ is a radio MSP, what is the probability that its radio beam
will pass the Earth (i.e. what is the beaming fraction of such pulsars)? A
period-dependent beaming fraction proportional to $P_{\rm spin}^{-0.5}$ has
been shown for normal, un-recycled pulsars \citep[see e.g.][]{ran93}. When
extrapolated to millisecond spin periods, this relation implies a $\sim
100$\% beaming fraction for MSPs.  However, \citet{kxl+98} find
observational evidence that this relationship does not apply directly to
MSPs and that the beaming fraction of MSPs is more like $50-90$\%. These
estimates are consistent with considerations of the MSP population in
47~Tuc. \citet{hge+05} used a deep {\it Chandra} observation of the cluster
to perform a population analysis which concluded that there are likely $\sim
25$ radio MSPs ($< 60$ at 95\% confidence) residing in the cluster,
independent of beaming.  Comparing this with optical and radio studies of
the MSP population by \citet{egh+03} and \citet{mdca04} respectively, who
both conclude that the total MSP population in 47~Tuc is $\sim 30$,
\citet{hge+05} conclude that the beaming fraction is $\gtrsim 37$\%.  While
these various studies suggest that the beaming fraction of MSPs is
significantly larger than for normal pulsars, and could be quite high, we
cannot strongly rule out beaming as the cause of our non-detection of \cal.

It is also possible, though we feel unlikely, that we have not seen \cal\
because it is in a compact binary orbit and/or is highly accelerated by a
binary companion.  Our luminosity limits are for a coherent search and do
not include these effects. However, given that we have employed acceleration
searches, which are generally sensitive in cases where the total integration
time $T_{\rm int}$ is less than roughly 1/10 of the orbital period $P_{\rm
orb}$, \cal\ would have to be in a $\lesssim 2$-hr orbit {\it and} fairly weak (or
eclipsed) not to have been detected in our searches of 15-min data sections.
The shortest orbital period of any known radio MSP is 1.6\,hr
\citep[][PSR~J0024$-$7204R in the GC 47~Tuc]{clf+00}.

In conclusion, given the caveats we have discussed, primarily pertaining to
extreme orbital or spin parameters, we believe that these searches were
sensitive enough to detect any nearby radio pulsar coincident with \cal,
assuming it is beamed towards the Earth.  We have checked for catalogued
radio sources in the FIRST, NVSS, and WENSS surveys and find no counterpart
to \cal.  Future, deeper observations detecting a radio, optical, or X-ray
pulsar wind nebula could address whether this source is a nearby radio
pulsar beamed away from Earth.  Alternate scenarios for \cal's nature, for
instance that it might be the first unhosted CCO (RFS07), should continue to
be investigated.

\begin{acknowledgements}
J.~W.~T.~H. acknowledges support from an NSERC postdoctoral fellowship and a
CSA supplement, tenured at the University of Amsterdam.  We would also like
to thank Gemma Janssen for help taking the observations.  The Westerbork
telescope is run by the Dutch radio astronomy institute ASTRON.

\end{acknowledgements}

\bibliographystyle{aa}
\bibliography{calvera}

%
%

\end{document}